\documentclass[Proceedings]{SciPost}
\usepackage[utf8]{inputenc}
\usepackage{booktabs}
\usepackage{graphicx}
\usepackage[caption=false]{subfig}
\usepackage{slashed}

\newcommand{\mysmall}{\scriptstyle \rm}
\newcommand{\emax}{\slashed E_{\mysmall max}}
\renewcommand{\Im}{\mathrm{Im}}

\begin{document}
\begin{flushright}
SI-HEP-2018-31,
QFET-2018-20
\end{flushright}

\bigskip
\begin{center}{\Large \textbf{ \boldmath
NLO prediction for the decays $\tau \to \ell \ell'\ell' \nu \bar \nu$ and $\mu \to e e e \nu \bar\nu$
  }}\end{center}

\begin{center}
Matteo Fael
\end{center}

\bigskip
\begin{center}
Theoretische Physik I, Universit\"at Siegen, Walter-Flex-Strasse 3, 57068 Siegen, Germany \\
fael@physik.uni-siegen.de
\end{center}

\definecolor{palegray}{gray}{0.95}
\begin{center}
\colorbox{palegray}{
  \begin{tabular}{rr}
  \begin{minipage}{0.05\textwidth}
    \includegraphics[width=8mm]{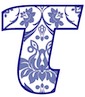}
  \end{minipage}
  &
  \begin{minipage}{0.82\textwidth}
    \begin{center}
    {\it Proceedings for the 15th International Workshop on Tau Lepton Physics,}\\
    {\it Amsterdam, The Netherlands, 24-28 September 2018} \\
    \href{https://scipost.org/SciPostPhysProc.1}{\small \sf scipost.org/SciPostPhysProc.Tau2018}\\
    \end{center}
  \end{minipage}
\end{tabular}
}
\end{center}


\section*{Abstract}
{
These proceedings review the differential decay rates and the branching ratios of the tau and muon decays $\tau \to \ell \ell' \ell' \nu \bar\nu$ (with $\ell,\ell'=\mu,e$) and $\mu \to e e e \nu \bar \nu$ in the Standard Model at NLO.
These five-body leptonic decays are a tool to study the Lorentz structure of weak interactions and to test lepton flavour universality. They are also a source of SM background to  searches for the lepton-flavour-violating decays $\mu \to e e e$ and $\tau \to \ell \ell' \ell'$.

Even if the shift in the branching ratios induced by radiative corrections turns out to be small and of order 1\% --- mainly due to a running effect of the fine structure constant --- locally in the phase space these corrections can reach the 5 - 10\% level, depending on the applied cuts.
We found for instance that in the phase space region where the neutrino energies are small, and the momenta of the three charged leptons have a similar signature as in $\mu \to eee$ and $\tau \to \ell \ell'\ell'$, the NLO corrections decrease the leading-order prediction by about 10 - 20\%.

}

\pagebreak

\section{Introduction}
\label{sec:intro}
About 35\% of the times the tau lepton decays only into electrons, muons and neutrinos. These leptonic decays of the tau (together with muon decay) constitute one of the more powerful tools to study precisely the structure of the weak interaction and possible contributions beyond the $V$--$A$ coupling of the Standard Model (SM) via the Michel parameters~\cite{Michel:1949qe,Bouchiat:1957zz,Kinoshita:1957zz,Kinoshita:1957zza}.
Michel parameters can be studied not only in three-body decays like $\tau \to \ell \nu \bar \nu$ and $\mu \to e \nu \bar \nu$, but also in muon and tau radiative modes~\cite{Eichenberger:1984gi,Fetscher:1993ki,Arbuzov:2016ywn,Shimizu:2017dpq},
\begin{align}
  \mu &\to e \nu \bar \nu \gamma, 
  \label{eqn:radmu} \\
  \tau &\to \ell \nu \bar \nu \gamma, \, \text{ with } \ell=\mu,e,
  \label{eqn:radtau}
\end{align} 
and in the rare five-body decays~\cite{Flores-Tlalpa:2015vga}
\begin{align}
\mu &\to e e e \nu \bar \nu, 
\label{eqn:raremu} \\
\tau &\to \ell \ell' \ell' \nu \bar \nu, \, \text{ with } \ell,\ell'=\mu,e.
\label{eqn:raretau} 
\end{align}
A study of five-body leptonic decays at \textsc{Belle} is ongoing with a data sample of about $0.91 \times 10^9$ $\tau^+\tau^-$ pairs~\cite{Sasaki:2017msf,Sasaki:2017unu}.
The measurement of the branching fractions and constraints on the Michel parameters will be presented soon.
Precise data on radiative and rare tau leptonic decays offer also the opportunity to probe the electromagnetic properties of the tau and they may allow to determine its anomalous magnetic moment~\cite{Laursen:1983sm,Eidelman:2016aih,Arroyo-Urena:2017ihp} which, in spite of its precise SM prediction~\cite{Eidelman:2007sb}, has never been measured.

Radiative and five-body decays of the tau and the muon constitute an important source of SM background to the searches for Charged Lepton Flavour Violation (CLFV) in $\mu \to e \gamma$, $\tau \to e \gamma$, $\mu \to e e e$ and $\tau \to \ell \ell'\ell'$ conversions.
These CLFV processes have been studied in the framework of the Standard Model Effective Field Theory (SMEFT)~\cite{Grzadkowski:2010es} to constraint in a model independent way possible sources of physics beyond the SM, violation of charged lepton flavour and lepton universality~\cite{Crivellin:2013hpa,Pruna:2014asa,Crivellin:2017rmk}.
Radiative decays are an important source of background to $\mu \to e \gamma$ and $\tau \to \ell \gamma$ searches, while the five-body muon decay~\eqref{eqn:raremu} is the main source of background to the search of the $\mu \to eee$ conversion (forbidden in the SM) at the \textsc{Mu3e} experiment at PSI~\cite{Blondel:2013ia}.
Indeed they are indistinguishable from the signal except for the energy carried out by neutrinos.
Secondly, these decays can be employed as a tool for calibration, normalization and quality check of the experiments~\cite{Adam:2013gfn,Kou:2018nap}.

The muon and tau leptonic decays can be predicted perturbatively in the SM since no low-energy QCD effect is involved. 
The Next-to-Leading Order (NLO) and Next-to-Next-to Leading Order (NNLO)  corrections to the muon lifetime were calculated in~\cite{Kinoshita:1958ru,vanRitbergen:1998yd,vanRitbergen:1998hn,vanRitbergen:1999fi}, together with the final state electron's energy spectrum~\cite{Anastasiou:2005pn} at NNLO, and the tree-level corrections induced by the $W$-boson propagator~\cite{Ferroglia:2013dga,Fael:2013pja}.
Concerning the radiative and the rare muon and tau decays, many tree-level calculations were presented in the last decades~\cite{Bardin:1972qq,Fishbane:1985xz,Dicus:1994dt,vanRitbergen:1999fi,Djilkibaev:2008jy,Flores-Tlalpa:2015vga,Arroyo-Urena:2017ihp}, but only recently a complete calculation of the NLO corrections to the muon and tau radiative decays~\cite{Fael:2015gua,Pruna:2017upz} and the rare muon five-body decays~\cite{Fael:2016yle,Pruna:2016spf,Ulrich:2017adq} were published.

These proceedings review the theoretical developments in the calculation of the NLO corrections for the muon decay~\eqref{eqn:raremu} in refs.~\cite{Fael:2016yle,Pruna:2016spf} and their impact on CLFV searches. Moreover, we also report the NLO prediction of the branching ratios for the tau five-body leptonic decays, that were omitted in the original publications~\cite{Fael:2016yle,Pruna:2016spf}. 

Section \ref{sec:methods} summarizes the methods employed in~\cite{Fael:2016yle} to calculate the NLO prediction of the differential decay rate and branching ratios, which are presented in section~\ref{sec:br}. Section~\ref{sec:clfv} is dedicated to discuss the importance of these radiative corrections in CLFV searches. Conclusions are drawn in section~\ref{sec:con}.

\section{Technical Ingredients}
\label{sec:methods}
In this section we describe the methods employed in~\cite{Fael:2016yle} for the calculation of the differential rate and the new development required in the evaluation of the five-body leptonic tau decays.
We adopted the Fermi $V$--$A$ effective theory of weak interactions:
\begin{equation}
  \mathcal{L} = 
  \mathcal{L}_\text{QED} 
  +\mathcal{L}_\text{QCD} 
  +\mathcal{L}_\text{Fermi}.
  \label{eqn:lfermi}
\end{equation}
The Fermi Lagrangian for the muon decays is
\begin{equation}
  \mathcal{L}_{\mysmall Fermi} =
  -\frac{4G_F}{\sqrt{2}}
  (\bar{\psi}_{\nu_\mu} \gamma^\alpha P_L \psi_\mu) \cdot
  (\bar{\psi}_e \gamma_\alpha P_L \psi_{\nu_e}) + \text{h.c.} \, ,
  \label{eqn:LFermi}
\end{equation}
while for the tau leptonic decays we used
\begin{equation}
  \mathcal{L}_{\mysmall Fermi} =
  -\frac{4G_F}{\sqrt{2}}
  (\bar{\psi}_{\nu_\tau} \gamma^\alpha P_L \psi_\tau) \cdot
  (\bar{\psi}_\ell \gamma_\alpha P_L \psi_{\nu_\ell}) + \text{h.c.} \, ,
  \label{eqn:LFermitau}
\end{equation}
with $\ell=\mu,e$ and where $\psi_\tau,\psi_\mu, \psi_e,\psi_{\nu_\tau},\psi_{\nu_\mu},\psi_{\nu_e}$ denote the fields of tau, muon, electron and their associated neutrinos, respectively; $P_L =(1-\gamma_5)/2$ denotes the left-hand projector.
Under this approximation tiny term of $\mathcal{O}(m_\mu^2/M_W^2) \sim 2 \times 10^{-6}$ and $\mathcal{O}(m_\tau^2/M_W^2) \sim 5 \times 10^{-4}$ due to the finite $W$-boson mass are neglected.

A Fierz rearrangement of the four-fermion interaction~\eqref{eqn:LFermi} and~\eqref{eqn:LFermitau} allows us to factorize the amplitudes of virtual and real corrections into the product of spinor chains depending either on the neutrino momenta or on the muon and electron ones (see Appendix~A.3 in~\cite{Fael:2016yle}).
Since the neutrino's phase space is integrated analytically, the residual phase space integration that must be performed numerically depends on a smaller number of integration variables.

The amplitudes for virtual corrections are reduced to tensor integrals and subsequently decomposed into their Lorentz-covariant structure by means of the algebra manipulation program \textsc{Form}~\cite{Kuipers:2012rf} and the \emph{Mathematica} package \textsc{FeynCalc}~\cite{Mertig:1990an,Shtabovenko:2016sxi}.
The matrix elements for one-loop and real emission diagrams are then exported as a Fortran code for the numerical integration.
Our code depends for the numerical evaluation of the tensor-coefficient functions on the \textsc{LooopTools}~\cite{Hahn:1998yk,vanOldenborgh:1989wn} and \textsc{Collier}~\cite{Denner:2016kdg} Fortran libraries, which can be both employed and compared.
The numerical integrations were performed with Monte Carlo methods by means of the \textsc{Vegas}~\cite{Lepage:1977sw} algorithm as implemented in the \textsc{Cuba} library~\cite{Hahn:2004fe}.

Ultraviolet (UV) divergences are regularized via dimensional regularization; UV-finite results are obtained by renormalizing the theory~\eqref{eqn:lfermi} in the on-shell scheme. A small photon mass $\lambda$ is introduced to regularize the infrared (IR) divergences, while the finite electron and muon masses regularize the collinear ones. 

The virtual corrections to the muon and tau five-body decays depends marginally on a non-perturbative contribution due the presence of the hadronic vacuum polarization in the off-shell photon propagator that converts into a $\ell^{'+}\ell^{'-}$ pair. 
This effect is not calculable at low energy in perturbative QCD but can be taken into account by expressing the hadronic vacuum polarization, $\Pi_{\mysmall had}(q^2)$, in terms of $e^+ e^- \to \text{hadrons}$ cross section data: 
\begin{equation}
  R_{\mysmall had}(s) = \sigma(e^+ e^- \to \, \text{hadrons})/
  \frac{4 \pi \alpha(s)^2}{3s}.
\end{equation}
The normalization factor $4 \pi \alpha(s)^2/(3s)$ is the tree-level cross section of $e^+e^- \to \mu^+ \mu^-$ in the limit $s \gg 4m_\mu^2$ --- note that $\sigma(e^+ e^- \to \, \text{hadrons})$ does not include initial state radiation or vacuum polarization corrections.
The optical theorem connects $R_{\mysmall had}(s)$ to the imaginary part of hadronic vacuum polarization:
\begin{equation}
  \Im \, \Pi_{\mysmall had} (s) =
  \frac{\alpha(s)}{3} 
  R_{\mysmall had} (s).
\end{equation}
The full vacuum polarization function can be then obtained by means of the dispersion relation. Our code employs the package \textsc{alphaQED}~\cite{Fred,Jegerlehner:2001ca,Jegerlehner:2006ju,Jegerlehner2011a} for the evaluation of the functions $R_{\mysmall had}$ and $\Pi_{\mysmall had} $.

In order to handle the IR divergences, we adopted in the original paper~\cite{Fael:2016yle} on the muon decay a phase-space slicing method. 
This method consists in the introduction of a small photon energy cut-off $\omega_0$ that divides the real emission contribution into a \emph{soft} and a \emph{hard} part. 
In the \emph{soft} part, containing the IR singularity, the photon's phase space is integrated analytically adopting the Eikonal approximation for the matrix element. The process-independent results was derived in~\cite{'tHooft:1978xw} (see also ref.~\cite{Denner:1991kt}). 
The \emph{hard} part, i.e.\ the contribution to the rate due to photons with energy greater than $\omega_0$, is integrated numerically and later merged with the result of the soft contribution. 
Since the result of this procedure is correct only up to $O(\omega_0/m_\mu)$ or $O(\omega_0/m_\tau)$, a precise prediction requires a rather small value of $\omega_0$.

For $\omega_0\to 0$, the numerical integration result grows like $\log \omega_0$, and therefore lots of CPU time is spent in the calculation of this known singular term that cancel out in the final result eventually. This issue turned out to be particularly severe in the calculation of the tau branching ratios. For this reason, for the five-body leptonic decays of the tau we employed the dipole subtraction method, originally developed in QCD~\cite{Catani:1996jh,Catani:1996vz} and later extended to QED in~\cite{Dittmaier:1999mb}. 
The idea is to carry out the phase-space integral of the real-emission matrix element, $|\mathcal{M}_\text{real}|^2$, without performing singular numerical integrations. 
To this end one subtracts and add an appropriate subtraction function, $|\mathcal{M}_\text{sub}|^2$, when integrating the $(n+1)$-particle phase space of the real emission:
\begin{equation}
  \int d \Phi_{n+1} |\mathcal{M}_\text{real}|^2 =
  \int d \Phi_{n+1} \left( |\mathcal{M}_\text{real}|^2 
   - |\mathcal{M}_\text{sub}|^2 \right)
   +\int d \Phi_{n+1} |\mathcal{M}_\text{sub}|^2.
   \label{eq:dipolesub}
\end{equation}
The subtracted term $|\mathcal{M}_\text{sub}|^2$ has the form:
\begin{equation}
  |\mathcal{M}_\text{sub}|^2 =
  - \sum_{f \neq f'} Q_f \sigma_f Q_{f'} \sigma_{f'}
  g_{f f'}^\text{(sub)} (p_f,p_{f'},k)
  |\mathcal{M}_\text{Born}|^2 ,
\end{equation}
where the sum runs over all charged fermions of the process, $Q_f$ and  $\sigma_f$ are the charge and the charge flow related to the fermion $f$ and $f'$, and the $g_{f f'}^\text{(sub)}$ are process-independent functions that possess the same asymptotic behaviour as $|\mathcal{M}_\text{real}|^2$ in the soft and collinear limit.
They depends on the  photon momentum $k$ and on the fermionic ones $p_f$ and $p_{f'}$. 
The momenta inserted inside the Born matrix element $|\mathcal{M}_\text{Born}|^2$, which depends on $n$ external momenta, are obtained by mapping the $n+1$ phase space into a $n$ particle phase space, respecting all mass-shell conditions. 
 In this way, the first term in~\eqref{eq:dipolesub} can be performed numerically without regulators while the second term, containing the IR singular contribution, is integrated analytically and then added to the contribution of virtual diagrams to cancel the IR poles. The part of the phase-space integral connected to the photon is process independent; its analytic integration was performed once and for all in~\cite{Dittmaier:1999mb}.

\section{\boldmath Branching ratios}
\label{sec:br}
\begin{table}[htb]
  \centering
  \begin{tabular}{lrrlr}
    \toprule
    & 
    $\quad \mathcal{B}_\mathrm{LO} \qquad$ &
    $\delta \mathcal{B}_\mathrm{lep}\qquad$ &
    $\quad \delta \mathcal{B}_\mathrm{had}$ &
    $\delta\cal B/\cal B$ \\
    \midrule
    $\tau \to eee\nu \bar \nu$ &
    $4.2488 \, (4) \times 10^{-5} $&
    $-4.2 \, (1) \times 10^{-8}$ &
    $-1.0 \times 10^{-9}$ &
    $-0.1 \% $\\
    $\tau \to \mu ee \nu \bar \nu$ &
    $1.989 \, (1) \times 10^{-5}$ &
    $4.4 \, (1) \times 10^{-8} $&
    $-6.6 \times 10^{-10}$ &
    $0.2 \%$\\
    $\tau \to e \mu \mu \nu \bar \nu$ &
    $1.2513 \,(6) \times 10^{-7}$&
    $2.70 \, (1) \times 10^{-9}$&
    $-3.6 \times 10^{-10}$ &
    $1.8 \%$ \\
    $\tau \to \mu \mu \mu \nu \bar \nu$ &
    $1.1837 \, (1) \times 10^{-7}$&
    $2.276 \, (2) \times 10^{-9}$ &
    $-3.5 \times 10^{-10} $ &
    $1.6 \% $\\
    \midrule
    $\mu \to eee\nu\bar\nu$ &
    $3.6054 \, (1) \times 10^{-5}$ &
    $-6.69 \,(5) \times 10^{-8}$ &
    $-1.8 \times 10^{-11}$ &
    $0.2\%$ \\
    \bottomrule
  \end{tabular}
  \caption{LO and NLO branching ratios of $\tau \to \ell \ell'\ell' \nu \bar \nu$ (with $\ell,\ell' = e,\mu$) and $\mu \to e e e \nu \bar \nu$. The NLO correction due to photons and leptons only is denoted by $\delta\mathcal{B}_\text{lep}$, while the non-perturbative contribution given by the hadronic vacuum polarization is denoted by $\delta\mathcal{B}_\text{had}$.
  The last column report the ratio between the NLO correction and the LO branching ratio. The uncertainties are the error due to numerical integration.}
\label{tab:BR}
\end{table}

The LO  branching ratios and the NLO corrections for the decays~\eqref{eqn:raremu} and~\eqref{eqn:raretau} are presented in Tab.~\ref{tab:BR}. 
All the branching ratios were computed keeping into account the full dependence on the lepton masses.
The second column shows the branching fraction at LO, while the third and the fourth columns  report separately the NLO contributions due to photons and leptons only (the dominant part) and the correction given by the hadronic vacuum polarization. 
The last column display the shift of the LO branching ratio induced by radiative corrections.
The uncertainties reported in Tab.~\ref{tab:BR} are the error of the numerical integration. 
We remind the reader that on top of the quoted uncertainties one must take into account for the tau lepton's decays also the error due to the tau lifetime; at present this error corresponds to a fractional uncertainty $\delta\tau_\tau/\tau_\tau = 1.7 \times 10^{-3}$~\cite{Tanabashi:2018oca}, which is of the same order of magnitude as the NLO corrections for the first two modes in Tab.~\ref{tab:BR}. For the rare muon decay the error due to the lifetime is negligible.

The NLO corrections to the branching ratios are of order $0.1\%$ for the tau decays involving at least two electrons (the first two modes in Tab.~\ref{tab:BR}) and the five-body decay of the muon. Radiative corrections are one order of magnitude larger for the tau decays into at least two muons (the third and fourth modes in Tab.~\ref{tab:BR}).
Such difference is caused by the running of the fine structure constant $\alpha$. 
In the decays $\tau \to e \mu \mu \nu \bar{\nu}$ and   $\tau \to \mu \mu \mu \nu \bar{\nu}$ the off-shell photon that converts into $\mu^+\mu^-$ is forced to acquire an invariant mass of at least twice the muon mass and therefore the electron's contribution to the photon vacuum polarization generates the logarithmic enhancement $\frac{\alpha}{3\pi} \log (4 m_\mu^2/m_e^2)$, which can be reabsorbed into the redefinition of $\alpha$ by substituting $\alpha \to \alpha(4m_\mu^2)$.
Note indeed that the shift induced by the running of $\alpha$ is $2 \times \Delta \alpha(4m_\mu^2) = 1.2 \%$, of the same order as the NLO corrections for the two modes with at least two muons in the final state.
This does not contradict the Kinoshita-Lee-Nauenberg theorem~\cite{Kinoshita:1962ur,Lee:1964is}, which guarantees that radiative corrections are free from mass singularities except for those that can be reabsorbed into the running of coupling constant. 

The branching ratio of~\eqref{eqn:raremu} was measured long time ago by the \textsc{Sindrum} experiment~\cite{Bertl:1985mw},
\begin{equation}
  \mathcal{B}_\text{exp} (\mu^- \to e^+ e^- e^- \nu_\mu \bar{\nu}_e) =
  3.4 \, (4) \times 10^{-5},
\end{equation}
while for the tau five-body decays, the \textsc{Cleo} experiment measured~\cite{Alam:1995mt}
\begin{equation}
  \mathcal{B}_\text{exp} (\tau \to e e^- e^+ \nu \bar{\nu}) =
  2.8 \, (1.5) \times 10^{-5},
\end{equation}
and established  for $\tau \to \mu e e \nu \bar{\nu}$ the upper bound
\begin{equation}
  \mathcal{B}_\text{exp} (\tau \to \mu e^- e^+ \nu \bar{\nu}) <
  3.2 \times 10^{-5} \text{ at } 90\% \text{ C.L.} \,.
\end{equation}
All available experimental measurements are in good agreement with the results in Tab.~\ref{tab:BR}. 
The \textsc{Belle} experiment is expected to present soon new measurements of the branching fractions for $\tau \to e e e \nu \bar{\nu}$ and $\tau \to \mu e e \nu \bar{\nu}$, a to report first upper bounds for the modes $\tau \to e \mu \mu \nu \bar{\nu}$ and   $\tau \to \mu \mu \mu \nu \bar{\nu}$~\cite{Sasaki:2017msf,Sasaki:2017unu,yifan}.

\section{\boldmath Impact on CLFV Searches.}
\label{sec:clfv}
\begin{figure*}[ht]
  \centering
  \subfloat[][]{\includegraphics[width=0.5\textwidth]{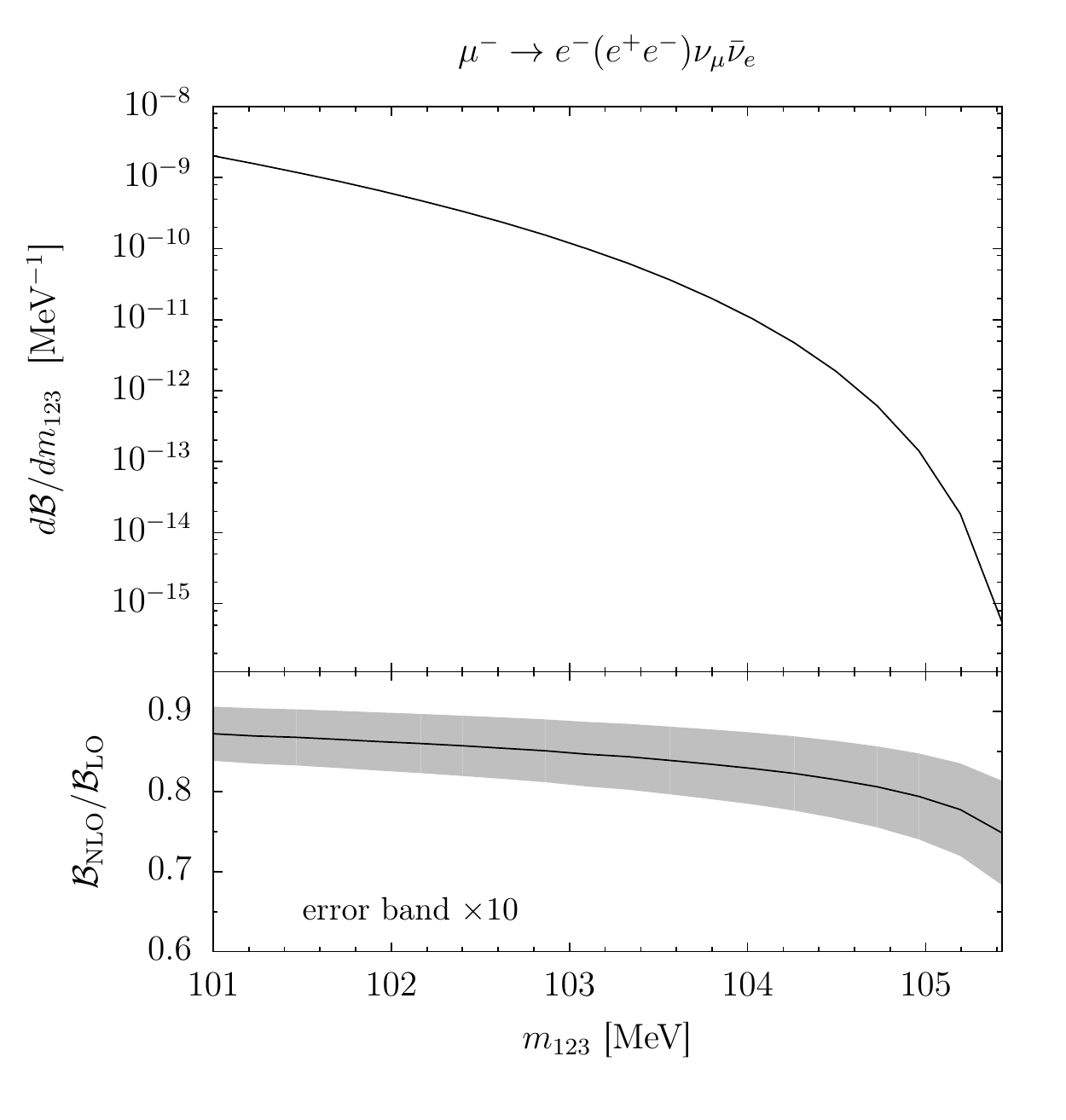}\label{fig:BRminv}}
  \subfloat[][]{ \includegraphics[width=0.5\textwidth]{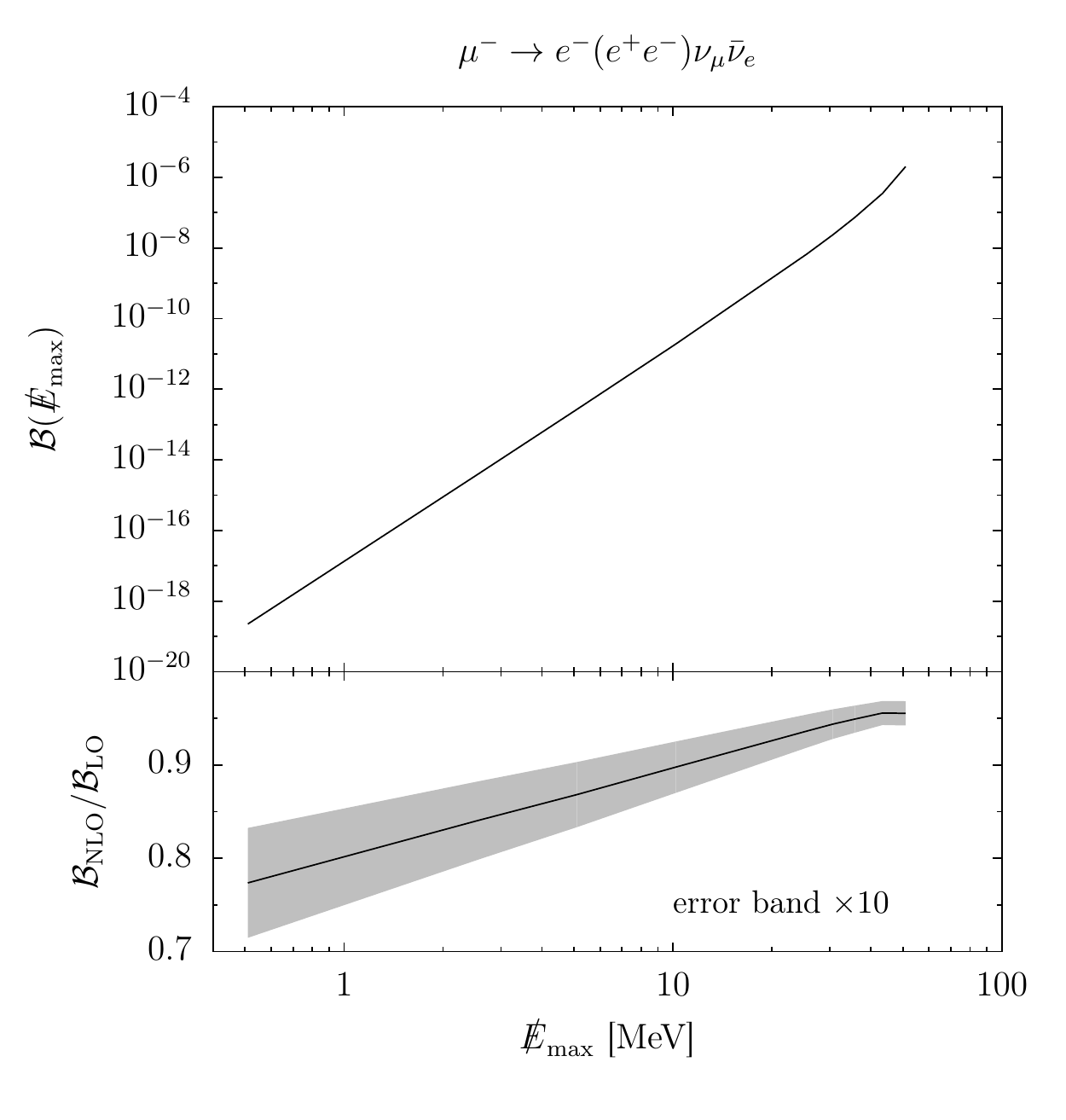}\label{fig:BRemax}
}
  \caption{The branching ratio of $\mu \to eee \nu \bar \nu$ at NLO as a function of the invariant mass of the three electrons $m_{123}$ (left) and the cut on the invisible energy $\slashed E_{\rm max}$ (right).
The ratio between the NLO and LO predictions is depicted in the lower part of each panel. 
The error band (magnified 10 times) represents the assigned theoretical error.
Figures taken from~\cite{Fael:2016yle}.}
  \label{fig:brall}
  \end{figure*}

Branching ratios are protected from large logarithmic corrections by the Kinoshita-Lee-Nauenberg theorem. 
However selection cuts on the final state can enhance the role of radiative corrections even up to 10\%.
As an example, we discuss here the size of these corrections in the specific final-state configuration of~\eqref{eqn:raremu} where the neutrino missing energy ($\slashed E$) is very small and the visible energy ($E_\text{vis}$) is close to $m_\mu$.
This region is particularly important for the \textsc{Mu3e} experiment.
Indeed, in this phase-space point the muon decay~\eqref{eqn:raremu} becomes a source of time- and space-correlated background for the CLFV three-body decay $\mu \to eee$.

Figure~\ref{fig:BRminv} displays  the normalized NLO differential rate as function of the three-electron invariant mass $m_{123}$, close to the end point $m_{123} = m_\mu$.
The local $K$-factor is shown in the lower part. The rate, evaluated at fixed value of $m_{123}$, is fully inclusive in the bremsstrahlung photon contribution.
%

Figure~\ref{fig:BRemax} shows the branching ratio $\mathcal{B}_{\mysmall NLO}(\emax)$ versus the cut on the missing energy (upper panel) and its relative magnitude with respect to the LO prediction (lower panel).
The branching ratio 
$\mathcal{B} (\slashed E_{\rm max})$ 
is calculated with a cut on the missing energy 
\begin{equation}
  \slashed E = m_\mu-E_\text{vis} \le \emax.
  \label{eqn:emax}
\end{equation}
Both distributions in figure~\ref{fig:brall} evince that radiative corrections decrease the LO prediction by about 10 - 20\%, depending on the cut applied on the missing energy. 
Hence, the background events for $\mu \to eee$ due to the decay~\eqref{eqn:raremu} are fewer than what is expected from a tree-level calculation.

The authors of ref.~\cite{Djilkibaev:2008jy} fit the LO missing energy spectum at the end point with the ansatz: 
\begin{equation}
  \mathcal{B}(\emax) = \kappa \left( \frac{\emax}{m_e} \right)^6,
  \mbox{ with } \kappa^\text{LO}=2.99 \times 10^{-19}.
  \label{eqn:fit1}
\end{equation}
We performed a similar fit for the NLO branching ratios for $\emax =1, 2, \dots, 10 \, m_e$. Taking into account the numerical error of $\mathcal{B}(\emax)$, we obtained the following value for the constant $\kappa$ at NLO accuracy:
\begin{equation}
  \kappa^\text{NLO}=2.5117 \,(6) \times 10^{-19}.
\end{equation}
The exponent of $\emax$ is assumed to be fixed in~\cite{Djilkibaev:2008jy}; relaxing such constraint and assuming also the exponent $\gamma$ to be a free parameter, i.e.\  
\begin{equation}
  \mathcal{B}(\emax) = \kappa' \left( \frac{\emax}{m_e} \right)^\gamma,
  \label{eqn:fit2}
\end{equation}
we obtain $ \kappa^{'\text{NLO}}=2.217\,(2) \times 10^{-19}$ and $\gamma^\text{NLO} = 6.0768 \,(4)$. 
Our ansatz~\eqref{eqn:fit2} is equivalent to a linear fit in a double logarithmic scale, $\ln \mathcal{B} = \ln \kappa' + \gamma \ln (\emax/m_e)$, while~\eqref{eqn:fit1} represents a straight line with fixed slope: $\ln \mathcal{B} = \ln \kappa + 6 \ln (\emax/m_e)$.

\section{Conclusion}
\label{sec:con}
We have reviewed the NLO predictions for the decay $\mu \to eee \nu \bar \nu$ and we have presented the NLO branching ratios for the tau five-body leptonic decays $\tau \to \ell \ell' \ell' \nu \bar \nu$, with $\ell,\ell'=e,\mu$.
Radiative corrections shift the branching ratio of about 0.1\%, for the decay modes with at least two electrons, and 1\% for the modes with at least two muons. 
These corrections are small because of cancellation of mass singularities in inclusive observables. The only logarithmic enhancement appearing in the modes $\tau \to e \mu \mu \nu \bar{\nu}$ and   $\tau \to \mu \mu \mu \nu \bar{\nu}$ is due too the running of the fine structure constant $\alpha$.

Detector acceptances and selection cuts can enlarge the magnitude of radiative corrections up to 10\% level.
In these proceedings we presented the case of $\mu \to eee \nu \bar \nu$ differential rate in the configuration where the total visible energy is close to the muon mass. In this corner of the phase space --- of particular importance for the \textsc{Mu3e} experiment since the decay mimics the CLFV decay $\mu \to eee$ --- these QED radiative corrections decrease the LO prediction by about 10 - 20 \%.

\section*{Acknowledgements}
This review is based on the work presented in~\cite{Fael:2016yle} and I would like to thank my collaborator C.\ Greub.
I would like to thank also G.\ M.\ Pruna, A.\ Signer and Y.\ Ulrich for valuable discussion and for sharing their results for the tau five-body leptonic decays, which are in good agreement with the results in Tab.~\ref{tab:BR}.
This work was supported by Swiss National Science Foundation and by DFG through the Research Unit FOR 1873 ``Quark Flavour Physics and Effective Field Theories''.

\bibliography{BIB}

\nolinenumbers

\end{document}